\documentclass[a4paper,11pt]{article}
\usepackage[hmarginratio=3:4,a4paper,tmargin=3truecm,bmargin=3truecm,rmargin=2.5truecm,
lmargin=2.5truecm,twoside,verbose=true]{geometry}
\usepackage{dsfont}
\renewenvironment{thebibliography}[1]
         {\section*{References}\frenchspacing\small
          \begin{list}{[\arabic{enumi}]}
         {\usecounter{enumi}\parsep=2pt\topsep 0pt
         \settowidth{\labelwidth}{[#1]}
         \leftmargin=\labelwidth\advance\leftmargin\labelsep
         \rightmargin=0pt\itemsep=1pt\sloppy}}{\end{list}}

\usepackage[english]{babel}
\usepackage[dvipsnames]{xcolor}
\usepackage{verbatim,ulem} 
\usepackage{pdfsync}
\usepackage{cancel,graphicx}
\usepackage{hyperref}
\usepackage{amsmath,amssymb,slashed,amsthm}
\usepackage{footmisc}

\newcommand{\aunk}{\mathcal{A}}
\newcommand{\A}{\mathcal{A}}
\newcommand{\B}{\mathcal{B}}
\newcommand{\I}{\mathbb{I}}
\newcommand{\V}{\mathcal{V}}
\newcommand{\HH}{\mathcal{H}}
\newcommand{\hh}{\mathcal{H}}

\newcommand{\R}{\mathbb{R}}
\newcommand{\C}{\mathbb{C}}
\newcommand{\Z}{\mathbb{Z}}

\newcommand{\N}{\mathbb{N}}
\newcommand{\inner}[1]{\left<#1\right>}
\newcommand{\sa}{{\cal S(\A)}}
\newcommand{\pa}{{\cal P(\A)}}

\newcommand{\sad}{{{\cal S}(\A_2)}}
\newcommand{\pad}{{{\cal P}(\A_2)}}

\newcommand{\zp}{{z_\Psi}}
\newcommand{\zpp}{{z_{\Psi'}}}
\newcommand{\of}{{\omega_{\phi}}}
\newcommand{\ofp}{{\omega_{\phi'}}}

\newcommand\caA{{\mathcal A}}
\newcommand\caH{{\mathcal H}}
\newcommand\bbbone{\mathbb{I}}

\newcommand\ub{{\mathcal B}_D}

\newcommand\dlde{ {\cal{D}}_{L^2} }

\newcommand\caS{{\mathcal S}}

\newcommand\gR{{\mathbb R}}
\def\gC{{\mathbb C}}

\def\gR{{\mathbb R}}

\newcommand\gN{{\mathbb N}}

\newcommand\algA{{\mathbf A}}

\newcommand\del{{\partial}}
\newcommand\delbar{{{\bar{\partial}}}}

\DeclareMathOperator{\tr}{Tr}

\newcommand\op{{\omega_\Psi}}

\newcommand{\norm}[1]{\left\lVert#1\right\rVert}
\newcommand{\abs}[1]{\lvert#1\rvert}

\newtheorem{Theorem}{Theorem}[section]

\newtheorem{Proposition}[Theorem]{Proposition}
 \newtheorem{Lemma}[Theorem]{Lemma}

\newtheorem{Remark}[Theorem]{Remark}
 \newtheorem{Definition}[Theorem]{Definition}

\numberwithin{equation}{section}

\begin{document}
% %%%%%%%%%%%%%%%%%% Title page for ArXiv %%%%%%%%%%%%%%%%%%%%%%%%%%%%%%
\title{The spectral distance in the Moyal plane}
\author{Eric Cagnache$^a$, Francesco D'Andrea$^b$, Pierre Martinetti$^{c,d}$ and Jean-Christophe Wallet$^a$}
\date{}
\maketitle
\vspace*{-1cm}
\begin{center}
\textit{$^a$Laboratoire de Physique Th\'eorique, B\^at.\ 210\\
    Universit\'e Paris-Sud 11,  91405 Orsay Cedex, France\\
    e-mail: \texttt{eric.cagnache@th.u-psud.fr}, 
\texttt{jean-christophe.wallet@th.u-psud.fr}}\\[1ex]
\textit{$^b$D\'epartement de Math\'ematiques, Universit\'e Catholique de Louvain\\
Chemin du cyclotron 2, 1348 Louvain-La-Neuve, Belgium\\
    e-mail: \texttt{francesco.dandrea@uclouvain.be}}\\[1ex]
\textit{$^c$Institut f\"ur Theoretische Physik, Georg-August Universit\"at\\
Friedrich-Hund-Platz 1, 37077 G\"ottingen, Germany}\\[1ex]
\textit{$^d$Courant Center  ``Higher Order Structures in Mathematics'',
  Universit\"at G\"ottingen\\
Busenstr. 3-5, 37073 G\"ottingen, Germany\\
    e-mail: \texttt{martinetti@theorie.physik.uni-goettingen.de}}
\\
\end{center}%
\vskip 2cm
% %%%%%%%%%%%%%%%%%% End of title page for ArXiv   %%%%%%%%%%%

\begin{abstract}\noindent
We study the noncommutative geometry of the Moyal plane from a metric
point of view. Starting from a non compact spectral triple based on
the Moyal deformation $\A$ of the algebra  of Schwartz functions on
$\R^2$,  we explicitly compute Connes' spectral distance between the 
pure states of $\A$ corresponding to eigenfunctions of the quantum harmonic
oscillator. For other pure states, we provide a lower bound to the
spectral distance, and show that the latest is not always
finite. As a consequence, we show that the spectral triple \cite{Gayral:2004rc} is not
a spectral metric space in the sense of \cite{Bellissard:2010fk}. This
motivates the study of truncations of the spectral triple, based on
 $M_n(\C)$  with arbitrary $n\in\N$, which turn out to be compact quantum metric
spaces in the sense of Rieffel. Finally the distance is explicitly
computed for $n=2$.
\end{abstract}

\pagebreak

\section{Introduction}

Mainly motivated by quantum mechanics, where physical quantities are no longer
functions on a manifold as in classical mechanics but elements of a 
noncommutative operator algebra, Noncommutative Geometry \cite{Con94} aims at descri\-bing
``spaces'' in terms of algebras $\A$ rather than as sets of points.
Many examples of noncommutative spaces are known (e.g.~almost commutative manifolds \cite{ISS}, fuzzy spaces \cite{Madore:1997vo}, deformations
for actions of $\R^n$ \cite{Rieffel:1990xp},
Drinfel'd-Jimbo~\cite{Dri86,Jim85},  Connes-Landi \cite{Connes:2001pi}
and Connes-Dubois-Violette~\cite{Connes:2002xr} deformations), but little work has been done
regarding their metric aspect. Let us recall that in Connes theory a natural  distance \cite{CL92} on the state space $\sa$ of $\A$  is obtained via the construction
of a \emph{spectral triple} (cf.~Sec.~\ref{spectraltriple}), the latest  providing a noncommutative genera\-lization
of many tools of differential geometry \cite{Con08}. In this paper we
focus on the associated distance, whose definition is recalled in Def.~\ref{spectraldist},  from now on called
the \emph{spectral distance}. 

On a (finite dimensional, complete) Riemannian spin manifold
$M$, the spectral distance $d$ between pure states of the commutative algebra $C_0^\infty(M)$ of  smooth functions vanishing at infinity coincides with the geodesic distance between the corresponding
points, while $d$ between non-pure states coincides with the Wasserstein distance of order $1$ between the correspon\-ding probability
distributions \cite{DM09}. In the noncommutative case, the meaning of
the spectral distance is still obs\-cure, essentially due to a lack of
examples: $d$ has been explicitly computed only for finite dimensional
noncommutative algebras (e.g.~\cite{IKM01,bls}) and almost commutative
geometries \cite{cc,kk} where it exhibits some interesting links with
the Carnot-Carath{\'e}odory distance in sub-Riemannian geometry
\cite{cs}. To shed more light on the spectral distance in a noncommutative framework, a natural idea is to investigate the
motion of a quantum particle along a line which, from a mathematical point of view,
amounts to studying the well known Moyal plane. An associated spectral triple has
been proposed in \cite{Gayral:2004rc}, built around the algebra $\A$ of Schwartz functions on $\R^2$ equipped with the Moyal product $\star$. 
Although this spectral triple is an isospectral deformation of the canonical spectral triple of the Euclidean plane,  the spectral distance $d$  on the Moyal plane does not appear to be a deformation of the Euclidean distance on $\R^2$. It has rather a quantum mechanics interpretation. Indeed the pure states $\pa$ of $\A$ correspond to Wigner  transition eigenfunctions of the quantum harmonic oscillator, and we show in this paper (Proposition \ref{zetheorem})
that the eigenstates of the quantum harmonic oscillator form a
$1$-dimensional lattice of the Moyal plane, with distance between two
subsequent energy le\-vels $E_{m-1}, E_m$ proportional to $m^{-\frac
  12}$.  We also point out some states at infinite distance from one
another, meaning that the topology induced by $d$ on $\sa$ is
\emph{not} the weak$^*$ topology. Therefore the spectral triple
for the Moyal plane introduced in \cite{Gayral:2004rc}  is
not a \emph{spectral metric space} \cite{Lat05, Bellissard:2010fk}.
 This leads us to study truncations of the Moyal spectral triple, which turn out to be \emph{compact quantum metric spaces}
in the sense of Rieffel \cite{Rie99,Rie03}.

The paper is organized as follows. In section 2 we recall some basics on the Moyal product, the associated spectral triple, the pure states of the Moyal algebra and we establish several technical results. In section 3, we explicitly compute the spectral distance between the eigenstates of the quantum harmonic oscillator%as well as on sub-spheres of $\pa$
. We also derive some bounds for the dis\-tance between any pure
states and show that the induced topology is not the weak$^*$ topology. In section \ref{sec:4}  we introduce a spectral triple on $M_n(\C)$
obtained by truncating the Moyal spectral triple. We explicitly compute the
associated spectral distance for $n=2$, and for $n\geq 2$ we prove that
the truncation gives a compact quantum metric space (Proposition \ref{quantumetrictheo}). This is an interesting result since the ``natural''
spectral triple on $M_n(\C)$ studied in \cite{IKM01} did not induce a quantum metric space.
In Section \ref{sec:5} we draw conclusions and illustrate
open problems.

\section{Moyal non compact spin geometries}
\subsection{Moyal product, matrix basis and relevant algebras of tempered\\ distribution on $\mathbb{R}^2$}

The main properties of the Moyal machinery can be found  e.g.~in \cite{Gracia-Bondia:1987kw,Bondia:1988qv} to which we refer for more details. Various related algebras have appeared in the literature, some of which will be recalled below. An extension of Connes real spectral triple to the non-compact case, to which we will refer heavily throughout this paper, has been carried out in \cite{Gayral:2004rc}; the corresponding action functionals and spectral actions have also been considered in \cite{Moyalspec}. Constructions of various derivation based differential calculi on Moyal algebras have been carried out in \cite{Cagn,WAL1,degour} together with applications to the construction of Yang-Mills-Higgs models on non-commutative Moyal spaces. In this work, we will only consider the 2-dimensional case. \par 
Let ${\cal{S}}= {\cal{S}}({\mathbb{R}}^2)$
be the space of complex Schwartz functions on ${\mathbb{R}}^2$ and
${\cal{S}}^\prime ={\cal{S}}^\prime({\mathbb{R}}^2)$ its topological
dual. Let $\theta>0$ be a fixed positive real parameter.
\begin{Proposition}\cite{Moyal,Groene} 
The  associative  Moyal $\star-$product is defined for all $a, b$ in ${\cal{S}}$ by \linebreak $\star:{\cal{S}}\times{\cal{S}}\to{\cal{S}}$
\begin{gather}
\label{eq:moyal0}
(a\star b)(x)=\frac{1}{(\pi\theta)^2}\int d^2y\,d^2z\ a(x+y)b(x+z)e^{-i\,2y\,\Theta^{-1}z},
\\
\text{ where }\; y\,\Theta^{-1}z\equiv y^\mu \Theta^{-1}_{\mu\nu}z^\nu,\ 
 \Theta_{\mu\nu}=\theta\begin{pmatrix} 0&1 \\ -1& 0 \end{pmatrix}.
\notag
\end{gather}
The complex conjugation is an involution for the $\star$-product; the integral is a faithful trace, 
\begin{equation*}
\int d^2x\ (a\star b)(x)=\int d^2x\ (b\star a)(x)=\int d^2x\ a(x)b(x);
\end{equation*}
the Leibniz rule is satisfied: $\partial_\mu(a\star b)=\partial_\mu a\star b+a\star\partial_\mu b$, $\forall a,b\in{\cal{S}}$.
\end{Proposition}
\noindent  In all the paper, we denote 
\begin{equation}
  \label{eq:18}
  {\cal{A}}:=({\cal{S}},\star)
\end{equation} the non-unital involutive algebra of Schwartz functions equipped with the Moyal product.

Our analysis will use the matrix basis whose relevant properties are summarized below.
\begin{Proposition}\label{matrix-basis}\cite{Gracia-Bondia:1987kw}
The matrix basis $\{f_{mn}\}_{m,n\in\mathbb{N}}\subset{\cal{S}}\subset L^2({\mathbb{R}}^2)$
is the family of  Wigner transition eigenfunctions of the 1-dimensional harmonic oscillator,
$$
f_{mn}={\frac{1}{(\theta^{m+n}m!n!)^{1/2}}}\;{\bar{z}}^{\star m}\star f_{00}\star z^{\star n}
$$
where $f_{00}=2e^{-2H/\theta}$,  $H={\frac{1}{2}}(x_1^2+x_2^2),$
$X^{\star n}= \underset{n\text{ times}}{\underbrace{X\star X\star...\star X}}$ and
\begin{equation}
\label{defz}
{\bar{z}}={\tfrac{1}{{\sqrt{2}}}}(x_1-ix_2), \qquad z={\tfrac{1}{{\sqrt{2}}}}(x_1+ix_2).
\end{equation}
\par 
\noindent i) Writing $\langle.,.\rangle$  the inner product{\footnote{Note the change of convention with respect to \cite{Gayral:2004rc} in which $\langle .,.\rangle$ denotes the inner product on $L^2(\R^2)$ divided by $\pi\theta.$}}
 on $L^2({\mathbb{R}}^2)$, one has the relations
$$
f_{mn}\star f_{pq}=\delta_{np}f_{mq},\qquad f_{mn}^*=f_{nm},\qquad \langle f_{mn},f_{kl} \rangle=(2\pi\theta)\delta_{mk}\delta_{nl}.
$$
ii) $\A$ is a Fr\'echet pre-$C^*$-algebra, with seminorms\footnote{With standard multi-indices notation
 $x^\alpha D^\beta:= x_1^{\beta_1}x_2^{\beta_2}\partial_1^{\alpha_1}\partial_2^{\alpha_2}$.} $\rho_{\alpha\beta}(a)=\sup_{x\in\gR^2} \abs{x^\beta D^\alpha a},$
isomorphic to the Fr\'echet algebra of rapid decay matrices  $(a_{mn})_{m,n\in{\mathbb{N}}}$  with seminorms
\begin{equation}
\rho_k^2(a) := \sum_{m,n\in\gN}\theta^{2k}(m+\tfrac 12)^k(n+\tfrac 12)^k|a_{mn}|^2.\label{eq:57}
\end{equation}
The isomorphism is given by $(a_{mn})\rightarrow \sum_{m,n}a_{mn}f_{mn}\in{\cal{S}}$, with inverse
\begin{equation*}
a\in{\cal{S}}\rightarrow  a_{mn}={\frac{1}{2\pi\theta}} \langle f_{mn},
a \rangle= {\frac{1}{2\pi\theta}}\int d^2x\,
a(x)f_{mn}^*(x)= {\frac{1}{2\pi\theta}}\int d^2x\, a(x)f_{nm}(x).
\end{equation*}
\end{Proposition}
\noindent 
The matrix basis diagonalizes the Hamiltonian of the harmonic oscillator,
$$
 H \star f_{mn} = \theta (2m +1) f_{mn},\; \qquad f_{mn} \star H = \theta (2n +1) f_{mn},
$$
and is also called the \emph{twisted Hermite
  basis} (an explicit decomposition on the Hermite functions $h_m$ is provided by the Wigner
operator $W(f_{mn}) = h_m\otimes h_n$, see \cite{Gracia-Bondia:1987kw}\footnote{The authors there use the convention $\theta = 2$.} and references therein). 
In particular  the 
$f_{mm}$, $m\in\N$, are the eigenstates of the $m^{\text{th}}$
energy level of the harmonic oscillator, and  in Proposition
\ref{zetheorem} below we compute the spectral distance between any two
of them. In a different context, the authors of \cite{Bahns:2010fk} and \cite{giacmoya} have used $H$
as (the square of) a distance-operator, yielding a notion of
quantized-distance in the Moyal plane which is distinct from the spectral distance. 
The link between these two approaches will be the object of a future work \cite{mm}.
For subsequent computations, let us write the derivatives of
the the $f_{mn}$, that are obtained by easy calculation.
\begin{Proposition}\label{relationcalcul}
Define $\partial={\frac{1}{{\sqrt{2}}}}(\partial_1-i\partial_2)$, ${\bar{\partial}}={\frac{1}{{\sqrt{2}}}}(\partial_1+i\partial_2)$.  For any $m,n\in{\mathbb{N}}$
\begin{align*}
\partial f_{mn}={\sqrt{{\frac{n}{\theta}}}}f_{m,n-1}-{\sqrt{{\frac{m+1}{\theta}}}}f_{m+1,n};\ {\bar{\partial}} f_{mn}={\sqrt{{\frac{m}{\theta}}}}f_{m-1,n}-{\sqrt{{\frac{n+1}{\theta}}}}f_{m,n+1}.\label{derivfmn}
\end{align*}
\end{Proposition}

The $\star$ product \eqref{eq:moyal0} is extended to
  spaces larger than ${\cal{S}}$, obtained by completing the Schwartz
  algebra 
with respect to the norm
\begin{equation*}
||a ||_{s,t}^2=\sum\nolimits_{m,n}\theta^{s+t}(m+\tfrac{1}{2})^s(n+\tfrac{1}{2})^t|a_{mn}|^2
\end{equation*}
with $s,t\in{\mathbb{R}}$.
Calling  ${\cal{G}}_{s,t}$ the completion, one has that 
for any $a \!=\!\sum_{m,n}a_{mn}f_{mn}\in{\cal{G}}_{s,t}$, 
$b\!=\sum_{m,n}b_{mn}f_{mn}\in{\cal{G}}_{q,r}$ with $t+q\geq 0$, the sequence 
$c_{mn}=\sum_{p}a_{mp}b_{pn}$ converges in ${\cal{G}}_{s,r}$, thus defining a map $\star:{\cal{G}}_{s,t}\times{\cal{G}}_{q,r} \rightarrow {\cal{G}}_{s,r}.$ 
In particular ${\cal{G}}_{0,0} = L_2(\R^2)$ and one has the dense
inclusions
${\cal{S}}\subset{\cal{G}}_{s,t}\subset{\cal{S}}^\prime$ for any $s,t\in\R$. Using
continuity and duality of $\star$ (for more details, see
e.g.~\cite{Gracia-Bondia:1987kw,Bondia:1988qv}) one can define the star
product on various subspaces of ${\mathcal S'}$. The following
algebras in particular are relevant for the study of Moyal spaces.
\begin{Proposition}\cite{Gayral:2004rc} \label{algebras} 
Let $\dlde$ denote the space of square integrable smooth functions 
on $\gR^2$ having all their derivatives in $L^2({\mathbb{R}}^2)$, ${\cal{B}}$ the space of smooth functions on $\mathbb{R}^2$ that are bounded together with all their derivatives and 
\begin{equation*}
\algA_\theta=\{a\in{\cal{S}}^\prime\; /\; a\star b\in L^2({\mathbb{R}}^2),\ \forall b\in L^2({\mathbb{R}}^2)\} \;.
\end{equation*}
i) $\algA_\theta$ is a unital C*-algebra with operator norm 
\begin{equation}
||a||_\theta=\norm{L(a)}_{op} = \sup_{0\ne b\in L^2({\mathbb{R}}^2)}\left\{{\frac{||a\star b||_2}{||b||_2}} \right\} \quad \forall a\in \algA_\theta, \label{eq:12}
\end{equation}
where  $L(.)$ is the left multiplication operator and $||.||_2$ denotes the $L^2({\mathbb{R}}^2)$ norm. Moreover $\algA_\theta$
is isomorphic to the algebra of bounded operators on $L^2(\gR)$.

\bigskip

\noindent
ii) $(\dlde,\star)$ is a Fr\'echet sub-algebra of
$(\algA_\theta,\star)$, with Fr\'echet topology of
$L^2({\mathbb{R}}^2)$-convergence for all derivatives;
$({\cal{B}},\star)$ is a Fr\'echet sub-algebra of
$(\algA_\theta,\star)$, with topology given by the family of
  semi-norms $\{p_m\}_{m\in{\mathbb{N}}},\
  p_m(a):=\text{max}_{\abs{\alpha}<m}||\partial^\alpha(a)||_\infty$.

\bigskip

\noindent
iii) $(\dlde,\star)$ and $({\cal{B}},\star)$ are respectively non
unital and unital pre C*-algebras with respect to $\norm{.}_\theta$.
Denoting  by an over-bar their $C^*$-completions, one has the inclusions
\begin{equation}\label{eq:59}
\A\subset\bar{\cal D}_{L^2}\subset \bar{\cal B}\subset\algA_\theta.
\end{equation}
\end{Proposition}
%\begin{Remark}
\noindent The above algebras are related but not identical to the
maximal unitization  ${\cal{M}}=({\cal{M}}_L\cap{\cal{M}}_R,\star)$ of $\A$ underlying most of the studies on noncommutative field theories
and noncommutative gauge theories on Moyal spaces (see e.g.~\cite{grossew,riv,Cagn,WAL1,degour} and references therein),
where
\begin{align*}
{\cal{M}}_L:= \{a\in{\cal{S}}^\prime\ /\ a\star b\in{\cal{S}},\ \forall b\in{\cal{S}}\},\qquad {\cal{M}}_R:=\{a\in{\cal{S}}^\prime\ /\ b\star a\in{\cal{S}},\ \forall b\in{\cal{S}}\}.
\end{align*}
${\cal{M}} $ 
 is unsuitable here as it cannot be represented by bounded operators
 on the Hilbert space $L^2(\R^2)\otimes\C^2$ used in the non-compact spectral triple that we now recall. \par 

\subsection{Spectral triple for Moyal plane}
\label{spectraltriple}
A  \emph{unital} spectral triple \cite{Con94} is the datum of a unital
involutive algebra $A$ together with  a representation $\pi$ on an
Hilbert space $\hh$ and a selfadjoint not necessarily bounded operator
$D$ (called Dirac operator) on $\hh$, such that $[D,\pi(a)]$ is
bounded for any $a\in A$ and the resolvent of $D$ is compact. In the
non compact case, i.e.~for non-unital $A$, one asks instead
\cite{Connes:1995kx}  that the operators $\pi(a)(D - \lambda)^{-1}$  are compact for any $\lambda \notin \text{Sp } D$. A natural candidate-spectral-triple for the Moyal plane is the isospectral deformation \cite{Connes:2001pi} of the canonical spectral triple of the Euclidean plane $\gR^2$ built around the algebra of Schwartz functions; namely
\begin{equation}
({\cal{A}},{\cal{H}},D)\label{eq:13}
\end{equation}
with $\A$ defined in \eqref{eq:18},
\begin{equation*}
{\cal{H}}= L^2({\mathbb{R}}^2)\otimes {\mathbb{C}}^2\equiv{\cal{H}}_0\otimes{\mathbb{C}}^2
\end{equation*}
the Hilbert space of square integrable sections of the trivial spinor bundle ${\mathbb{S}}={\mathbb{R}}^2\times{\mathbb{C}}^2$ with inner product
\begin{equation*}
\langle \psi,\phi\rangle=\int(\psi_1^*\phi_1+\psi_2^*\phi_2)d^2x \quad\forall
\;
\psi = \binom{ \psi_1 }{ \psi_2 }
,\;\phi = \binom{ \phi_1 }{ \phi_2 }\; \in{\cal{H}},
\end{equation*}
and, using  Einstein convention of summing over repeated indices,
\begin{equation}
D=-i\sigma^\mu\del_\mu = -i{\sqrt{2}}\begin{pmatrix} 0&\delbar \\ \del& 0 \end{pmatrix}
\label{eq:14}
\end{equation}
where 
\begin{align*}
\sigma^1=\begin{pmatrix} 0&1 \\ 1& 0 \end{pmatrix},\qquad
\sigma^2=\begin{pmatrix} 0&i \\ -i& 0 \end{pmatrix}
\end{align*}
span an irreducible representation of the Clifford algebra  $C\ell_{2}(\C)=M_2(\mathbb{C})$. ${\cal{A}}$ acts faithfully on ${\cal H}$ via the represen\-ta\-tion $\pi(a)=L(a)\otimes\bbbone_2$,
namely
\begin{align}
\label{eq705}
\pi(a)\psi=(a\star\psi_1,a\star\psi_2) \quad \forall a\in\A, \psi\in\hh.
\end{align}
\eqref{eq:13}  does satisfy the property of a non-compact spectral triple: 
$\pi(a)$ is bounded  by (\ref{eq705}) and \eqref{eq:59} together
  with $\algA_\theta \simeq {\cal{B}}(L^2(\R))$. The boundedness of
$[D,\pi(a)]$ comes from $[D,\pi(a)]=-i L(\partial_\mu
a)\otimes\sigma^\mu$, see \eqref{eq:2}, combined with $L(\partial_\mu
a)\in{\cal{B}}(L^2(\R))$. $D$ is the usual Dirac operator
  on $\R^2$ so it is essentially selfadjoint.
Note that $D$ being formally selfadjoint on its domain $\dlde\otimes{\mathbb{C}}^2$ can be easily seen noticing that the adjoint of $\del$ is
  $\del^\dag=-\delbar$. The resolvent condition is more involved and we refer to \cite{Gayral:2004rc} for the details.

In order to establish the equivalence between (unital) commutative spectral triples and  (compact, oriented, without boundary) smooth manifolds,
one further asks five extra-conditions on $(\A, {\cal H}, D)$ \cite{Con08}
 (dimension, order one, regularity, orientability, finiteness), that can be completed by two other conditions (reality, Poincar\'e duality) 
in order to recover spin and Riemannian structures from purely algebraic data. This yields the definition of \emph{real} spectral triples \cite{Connes:1995kx}. Among these conditions,  several easily translate to the non-compact case (dimension, regularity, reality, first order condition), one still asks a formulation in the non-compact case
(Poincar\'e duality). For the remaining two (finiteness, orientability), an adaptation to the non-unital case has been proposed in \cite{Gayral:2004rc}, based on a preferred unitization   
$ {\cal{A}}_1:=({\cal{B}},\star)$ of ${\cal{A}}$. However, as can be checked in  Definition \ref{spectraldist} below, none of these conditions
enters the definition of the spectral distance. So, regarding the metric aspect of the Moyal plane, it is more natural  to  work with the spectral triple \eqref{eq:13} as we do in the following than with the unitization ${\cal A}_1$. We come back to this point in section 4 and in the conclusion.

\section{Spectral distance on the Moyal plane}

\subsection{Distance formula and pure states}

Let us begin with a technical precision. A state on a complex $C^*$-algebra $A$ is a positive linear map $A \rightarrow \gC$ of norm $1$, and strictly speaking this notion is reserved for $C^*$-algebras. However given a pre $C^*$-algebra $A$, a state on its $C^*$-completion $\bar A$ defines by restriction a unique
positive linear map of norm $1$ from $A$ to $\gC$, and conversely by continuity any state of $\bar A$ is uniquely determined by its restriction to $A$.
So it is also legitimate to talk about states for pre $C^*$-algebras.

The metric aspect of the noncommutative geometry $(\caA, \caH, D)$ introduced in (\ref{eq:13}) is fully encoded within the spectral distance defined as follows.
\begin{Definition}
\label{spectraldist}
The spectral distance between two states $\omega_1$ and $\omega_2$ of $ \aunk$ is
\begin{equation}
d(\varphi_1,\varphi_2)=\sup_{a\in\aunk}\big\{|\varphi_1(a)-\varphi_2(a)|;\ ||[D,\pi(a)] ||_{\text{op}}\le1\big\}\label{spectralformule}
\end{equation}
where $||.||_{\text{op}}$ is the operator norm for the representation of ${\cal{A}}$ in ${\cal{B}}({\cal{H}})$.
\end{Definition}
\noindent
One easily checks that \eqref{spectralformule} has all the
properties of a distance,  except that it may take the value $+\infty$.
Recall that in the commutative case, i.e.~$\A=C^\infty_0(M)$ for $M$ a complete Riemannian spin manifold, $\HH$ the space of spinors square-integrable with respect to the volume form associated to the Riemannian metric, and $D$ the Dirac operator of the Levi-Civita
connection, then pure states are evaluations at $x\in M$:
  \begin{equation}
    \label{eq:62}
    \delta_x(f) = f(x) \quad \forall f\in C^\infty_0(M),
  \end{equation}
  and $d(\delta_x, \delta_y)$ coincides with the geodesic distance on $M$.
Therefore it is appealing, in the noncommutative case, to consider the spectral distance
between pure states as a good equivalent 
of the geodesic distance between points.

However, on the deformed algebra $\A$ the evaluation~\eqref{eq:62} is no
longer a state, for
$(f^*\star f)(x)$
 has no reason to be positive. Having in mind Moyal spaces as a
 quantized version of Euclidean spaces, one could be tempted to look
 at the pure states of  $\A$ as a deformation of the pure states of $C^\infty_0(\R^2)$. A nice approach on how to obtain states for a Rieffel deformation of a $C^*$-algebra 
by deformation of states of the undeformed algebra has been developed
in \cite{wald}, but the ``purity'' of the state is not addressed
there. In fact, rather than using the $x$-representation \eqref{eq:moyal0}, pure states of $\A$
are more easily determined in the twisted Hermite basis
$\left\{f_{mn}\right\}_{m,n \in \N}$.

Let $e_n$ be the canonical orthonormal basis of $\ell^2(\N)$ and $\eta$ the natural
representation of rapid decay matrices, given by row by column multiplication.
By Prop.~\ref{matrix-basis} the isometry $L^2(\R^2)\to \ell^2(\N)\otimes\ell^2(\N)$ given
by
$$
U:f_{mn}\mapsto \sqrt{2\pi\theta}\,e_m\otimes e_n
$$
is an intertwiner between the left regular representation $L$ and the representation
$\eta\otimes 1$. One can see that $\eta$ is faithful and irreducible\footnote{It is irreducible
since its restriction to $M_\infty(\C)$ is irreducible.}, while $L$
is only faithful (it is the direct sum of infinitely many copies of $\eta$, 
in the same way as the left regular representation of $M_n(\C)$ with Hilbert-Schmidt
inner product is unitary equivalent to the direct sum of $n$ copies of the irreducible
representation). Therefore
$$
\A\simeq L(\A)\simeq \eta(\A) \;.
$$
Since the norm of any operator $a$ on $\ell^2(\N)$ coincides with the
norm of $a\otimes 1$ on $\ell^2(\N)\otimes\ell^2(\N)$,
the closure of 
$L(\A)$ in $\B(L^2(\R^2))$ is isometrically $*$-isomorphic to the closure
of $\eta(\A)$ in $\B(\ell^2(\N))$: we call $\bar{\A}$ this $C^*$-algebra.

Notice that
$$
\bar{\A}\supset\eta(\A)\supset M_\infty(\C):=\bigcup\nolimits_{k\geq 1}M_k(\C)
$$
where $M_k(\C)$ are identified with $a=((a_{mn}))\in\A$ such that
$a_{mn}=0$ if $m\geq k$ or $n\geq k$.
By \cite[II.8.2.2]{blackadar2006}
the closure of $M_\infty(\C)$ is the $C^*$-algebra $\mathbb{K}$ of
compact operators on $\ell^2(\N)$, proving that 
$\bar{\A}=\overline{\eta(\A)}\supset\mathbb{K}$.
On the other hand, $\eta$ maps any rapid decay matrix $a$ into a Hilbert-Schmidt
operator, since the norm $\rho_0(a)$ in \eqref{eq:57} is exactly the Hilbert-Schmidt norm
of $\eta(a)$. Hence $\eta(a)$ is a compact operator for any $a\in\A$ and
$\overline{\eta(\A)}\subset\mathbb{K}$, proving that $\bar{\A}\simeq\mathbb{K}$.

It is well known \cite[Cor.~10.4.4]{Kadison:1986zm} that all pure states
of $\mathbb{K}$ are vector states of the (unique) irreducible representation
on $\ell^2(\N)$. Hence,
\begin{Proposition}\label{purestates}
Any unit vector $\psi=\sqrt{2\pi\theta}\sum_m\psi_me_m\in\ell^2(\N)$, i.e.~with normalization $\sum_m|\psi_m|^2=\frac{1}{2\pi\theta}$,
defines a pure state $\omega_\psi$ of $\aunk$,
\begin{equation}\label{eq:19}
\omega_\psi (a) := 2\pi\theta \sum_{m,n\in{\mathbb{N}}} \psi^*_m\psi_n  a_{mn} .
\end{equation}
Moreover any pure state of $\aunk$ comes from such a unit vector. 
\end{Proposition}

\noindent
Notice that $\psi$ can be obtained from
\begin{equation}
v_n:=U^*(\psi\otimes e_n)=\sum\nolimits_{m\geq 0}\psi_m f_{mn} 
\label{eq:39}
\end{equation}
for any $n\in\N$, and we have $\omega_\psi (a)\equiv \left<v_n,L(a)v_n\right>$.

\subsection{Unit ball}
In order to compute the distance \eqref{spectralformule} on the Moyal plane, one needs to conveniently  charac\-terize the unit ball
\begin{equation}
\label{unitball}
  \ub:= \left\{ a\in\aunk , \, ||[D,\pi(a)] ||_{\text{op}}\le 1 \right\}.
\end{equation}

The first step is to determine the relation between the coefficients of $a$ in the matrix basis and those of $[D,\pi(a)]$. Noticing that $[\partial_\mu,L(a)]=L(\partial_\mu a)$ by Leibniz rule
one obtains
\begin{equation}
    \label{eq:2}
    [D,\pi(a)] = -i\sqrt{2}\left(
\begin{array}{cc} 0& L(\bar\partial a)  \\ L(\partial a) &0 
\end{array}\right).\end{equation}
Thus 
\begin{align}
\norm{[D,\pi(a)]}_{\text{op}}^2 &= \norm{[D,\pi(a)]^*[D,\pi(a)]}_{\text{op}} =
 2\norm{
\begin{pmatrix} L(\partial a)^*L(\partial a) & 0 \\ 0 & L (\bar\partial a)^*L(\bar\partial a) 
\end{pmatrix}}_{\text{op}} \notag\\
&= 2\max\left(
\norm{L(\partial a)^*L(\partial a)}_{\text{op}}, \norm{L(\bar\partial a)L(\bar\partial a)^*}_{\text{op}} 
\right) \notag\\
&= 
 \label{eq:3bis} 2\max\left(
\norm{\partial a}^2_{\theta}, \norm{\bar\partial a}^2_{\theta} \right)
\end{align} 
and the point is now to find the relation between the
coefficients of $a$ and those of $\del a, \delbar a$.
\begin{Proposition}
\label{relationalpha}
For any $a\in\aunk$, $a=\sum_{m,n}a_{mn}f_{mn}$, we denote
\begin{equation*}
\del a :=\sum_{m,n}\alpha_{mn}f_{mn},\quad \delbar a:= \sum_{m,n}{{\beta}}_{mn}f_{mn}.
\end{equation*}

\par 
i) The coefficients of $\del a$, $\delbar a$ as functions of the coefficients of $a$ are given by
\begin{align}
\alpha_{m+1,n}={\sqrt{{\frac{n+1}{\theta}}}}a_{m+1,n+1}-{\sqrt{{\frac{m+1}{\theta}}}}a_{m,n},\ \alpha_{0,n}={\sqrt{{\frac{n+1}{\theta}}}}a_{0,n+1}, 
\label{alphamn}
\end{align}
\begin{align}
{{\beta}}_{m,n+1}={\sqrt{{\frac{m+1}{\theta}}}}a_{m+1,n+1}-{\sqrt{{\frac{n+1}{\theta}}}}a_{m,n},\ {{\beta}}_{m,0}={\sqrt{{\frac{m+1}{\theta}}}}a_{m+1,0},
\label{betamn}
\end{align}
for all $m,n\in{\mathbb{N}}$.

ii) One has the inversion formula
$$
a_{p,q}=\delta_{p,q}a_{0,0}+\sqrt{\theta}\,\sum_{k=0}^{\min(p,q)}
\frac{\alpha_{p-k,q-k-1}+\beta_{p-k-1,q-k}}{\sqrt{p-k}+\sqrt{q-k}}, \forall p,q\in{\mathbb{N}}, p+q>0
$$
where by convention terms with negative indices do not appear in the sum.
\end{Proposition}
\begin{proof} i) is proved  by standard calculation using Proposition \ref{relationcalcul}. To prove ii) one combines \eqref{alphamn} and \eqref{betamn} to obtain
\begin{align}
a_{p+1,q+1}=
a_{p,q}+\sqrt{\theta}\,\frac{\alpha_{p+1,q}+\beta_{p,q+1}}{\sqrt{p+1}+\sqrt{q+1}}.\nonumber
\end{align}
This yields by induction
\begin{align}
a_{p,q}=
\begin{cases}
a_{0,q-p}+\sqrt{\theta}\,\sum_{k=0}^{p-1}
\frac{\alpha_{p-k,q-k-1}+\beta_{p-k-1,q-k}}{\sqrt{p-k}+\sqrt{q-k}} &\text{if}\;0\leq p<q \;,\\
a_{0,0}+\sqrt{\theta}\,\sum_{k=0}^{p-1}
\frac{\alpha_{p-k,p-k-1}+\beta_{p-k-1,p-k}}{2\sqrt{p-k}} &\text{if}\;0\leq p=q \;,\\
a_{p-q,0}+\sqrt{\theta}\,\sum_{k=0}^{q-1}
\frac{\alpha_{p-k,q-k-1}+\beta_{p-k-1,q-k}}{\sqrt{p-k}+\sqrt{q-k}} &\text{if}\;0\leq q<p \;,
\end{cases} \nonumber
\end{align}
and ii) follows from a further use of the second relations for $\alpha_{0,n}$ and $\beta_{0,n}$.
\end{proof}

The second step in order to characterize ${\cal B}_D$ would be to compute $\norm{[D,a]}$ as an explicit function $f(\alpha_{mn})$, then use Proposition \ref{relationalpha}
 to transfer the constraints $f (\alpha_{mn})\leq 1$ to the
 coefficients $a_{mn}$. However  we are dealing with infinite
 dimensional matrices and one may well not have such a function
 $f$. Here we simply exhibit some necessary constraints on the
 $a_{mn}$'s so that $a\in {\cal B}_D$. Quite remarkably these
 constraints are sufficient to explicitly compute,  in the next
 section,  the distance for interesting classes of pure states.
\begin{Lemma}\label{lemme2}
Let  $a\in \ub$. Then\par
i) For any unit vector $\varphi=\sum_{m,n}\varphi_{mn}f_{mn}\in{\cal{H}}_0$ one has
\begin{align}
\label{lemalphabeta}
\sum_{p}|\alpha_{mp}||\varphi_{pn}| \le{\frac{1}{2{\sqrt{\pi\theta}}}}\;\; and \;\; \sum_{p}|\beta_{mp}||\varphi_{pn}| \le{\frac{1}{2{\sqrt{\pi\theta}}}},\quad\forall m,n\in{\mathbb{N}}; \
\end{align}
\par
ii) $|\alpha_{mn}|\le{\frac{1}{{\sqrt{2}}}}$\; and \; $|\beta_{mn}|\le{\frac{1}{{\sqrt{2}}}}$,\quad $\forall m,n\in{\mathbb{N}}.$
\end{Lemma}
\begin{proof}
Let $\varphi$ be a unit vector in $\hh_0$, that is to say $|| \varphi||^2_2= 2\pi\theta \sum_{mn} \abs{\varphi_{mn}}^2=1.$
Using the matrix basis, a standard calculation yields
\begin{equation*}
||\del a\star\varphi ||_2^2=2\pi\theta\sum_{m,n}\Big|\sum_p\alpha_{mp}\varphi_{pn}\Big|^2.
\end{equation*}
 By (\ref{eq:3bis}) $a\in\ub\!$ implies $||\del a||_\theta\!\le\!\!{\frac{1}{{\sqrt{2}}}}$ hence, owing to the definition of $||\del a||_\theta$, 
\begin{equation*}
\sum_{m,n}\Big|\sum_p\alpha_{mp}\varphi_{pn}\Big|^2\le{\frac{1}{4\pi\theta}}.
\end{equation*}
This implies 
\begin{align}
\Big|\sum_p\alpha_{mp}\varphi_{pn}\Big|\le{\frac{1}{2{\sqrt{\pi\theta}}}}\quad  \forall m,n\in{\mathbb{N}} \label{PP1}
\end{align}
together with a similar relation stemming from $||\delbar a||_\theta$ with the $\alpha_{mn}$'s replaced by $\beta_{mn}$.
Now (\ref{PP1}) holds true for any unit $\varphi\in{\cal{H}}_0$  and
in particular, given a value $m\in \gN$, for ${\tilde{\varphi}}^m$ defined by
 $\tilde\varphi_{pn}^m := e^{-i\text{Arg}(\alpha_{mp})}\abs{\varphi_{pn}}$.
 This implies 
\begin{align}
\sum_{p}|\alpha_{mp}||\varphi_{pn}|\le{\frac{1}{2{\sqrt{\pi\theta}}}},\quad\forall n\in{\mathbb{N}} \label{PP2}
\end{align}
and the first part of (\ref{lemalphabeta}) by repeating the procedure for other values of $m$. The second part is obtained 
by similar considerations apply to the $\beta_{mn}$'s.  ii) is obtained by considering in (\ref{PP2}) the unit vector $\varphi = \frac{1}{\sqrt{2\pi\theta}} f_{mn}$. 
\end{proof}
\noindent  Notice that  by the triangle inequality \eqref{PP2}
implies (and thus is equivalent to) \eqref{PP1}.
\begin{Remark}
\label{autoadjoint}
When the algebra $\caA$ is unital  the supremum in the distance formula can be equivalently searched on the positive unit sphere\cite{IKM01} 
\begin{equation*}
 {\cal S}_D^+:= \left\{ a\in \caA^+, \, ||[D,\pi(a)] ||_{\text{op}} =1\right\} 
\end{equation*}
where $\caA^+$ denotes the positive elements of $\caA$.
When the algebra is not unital, the supremum can be searched on the selfadjoint elements of the unit ball
\begin{equation*}
   {\cal B}_D^{\text{sa}}:= \left\{ a=a^*\in \caA, \, ||[D,\pi(a)] ||_{\text{op}} \leq 1\right\}.
 \end{equation*}
In this paper we are interested in the spectral distance on the non
unital algebra $\A$, so we assume that $a$ is always selfadjoint,  which implies $(\partial a)^* = (\bar\partial a)$, that is to say $\beta_{mn} = \bar\alpha_{nm}$.
\end{Remark}

\subsection{Distance on the diagonal}

The set of diagonal elements of $\mathcal{A}$
(i.e.~the maxi\-mal abelian sub-algebra)  is the set of rapid decay sequences $\mathcal{S}(\mathbb{N})$, whose pure states $\omega_m, m\in\gN$
are defined in \eqref{eq:19} with $\psi =e_m$ (i.e.~from \eqref{eq:39}: $U^*(\psi\otimes e_q)=\frac{1}{\sqrt{2\pi\theta}}f_{mq}$), namely
$$\omega_m(a) = a_{mm}.$$
\noindent Since on $\mathcal{S}(\mathbb{N})$  the operator norm and the max norm coincide, $||a||_{\mathrm{op}}\equiv\max_{p,q}|a_{pq}|$,
one can expect the distance between any two $\omega_m , \omega_n$ to be easily computable.
\begin{Proposition}
\label{zetheorem}
The spectral distance between $\omega_m$, $\omega_n$, $n<m$, is
$$
d(\omega_m,\omega_n)={\sqrt{{\frac{\theta}{2}}}}\sum_{k=n+1}^m{\frac{1}{{\sqrt{k}}}}.
$$
\end{Proposition}
\begin{proof}
By  Proposition \ref{relationalpha}, 
\begin{align}
\alpha_{n+1,n}={\sqrt{{\frac{n+1}{\theta}}}}(a_{n+1,n+1}-a_{nn})={\sqrt{{\frac{n+1}{\theta}}}}(\omega_{n+1,n+1}(a)-\omega_{nn}(a)),\ \forall n\in{\mathbb{N}}\nonumber
\end{align}
so that, for any $a$ in the unit ball,  ii) of Lemma \ref{lemme2} yields 
\begin{equation}
|\omega_{n+1}(a)-\omega_{n}(a)|\le{\sqrt{{\frac{\theta}{2}}}}{\frac{1}{{\sqrt{n+1}}}},\quad \forall n\in{\mathbb{N}}.
\label{eq:15}
\end{equation}
Hence $d(\omega_{n+1}, \omega_n) \leq{\sqrt{{\frac{\theta}{2}}}}{\frac{1}{{\sqrt{n+1}}}}$ and by  a repeated use of the triangular inequality
\begin{equation}
d(\omega_m,\omega_n)\le{\sqrt{{\frac{\theta}{2}}}}\sum_{k=n+1}^m{\frac{1}{{\sqrt{k}}}},\quad \forall m,n\in{\mathbb{N}},\ n<m.
\label{eq:16}
\end{equation}
\par 
This upper bound is attained by any element in the unit ball which saturates \eqref{eq:15} between $m$ and $n$, that is to say such that
\begin{equation}
  \label{eq:13ter}
a_{p+1,p+1} - a_{pp} = \sqrt{\frac{\theta}{2}} \frac 1{\sqrt{p+1}}\quad \text{ for } n\leq p\leq m-1.
\end{equation}
For instance the element $a^{(m)}$ (diagonal in the matrix basis) with components
\begin{equation}
  \label{eq:10}
  a^{(m)}_{pq} = -\delta_{pq}\sqrt{\frac {\theta}2}\sum_{k= p}^m \frac
   {1}{\sqrt{k+1}} 
\end{equation}
(where an empty sum means zero) obviously satisfies \eqref{eq:13ter}. Moreover $\left\{a^{(m)}_{pq}\right\}_{p,q\in\gN}$ is a rapid decay sequence since
only finitely many elements are non-zero, hence $a^{(m)}\in\caA$. By i) of Proposition \ref{relationalpha}, one checks that all the $\alpha_{pq}$'s vanish except
$\alpha_{p+1,p} =  \frac 1{\sqrt{2}} \text{ for } 0\leq p\leq m.$ Therefore for any $\psi\in\hh_0$,
\begin{align*}
  \norm{\rule{0pt}{10pt}\smash[t]{\partial a^{(m)} \star \psi}}_2^2 &= 2\pi\theta \sum_{p,q}\Big|\sum_{g} \alpha_{pg} \psi_{gq}\Big|^2
= 2\pi\theta \sum_{p,q}\big| \alpha_{p+1,p} \psi_{pq}\big|^2 \\
&\leq \pi\theta \sum_{p,q}\abs{\psi_{pq}}^2 
= \frac 12 \norm{\psi}^2,
\end{align*}
so that $\norm{\partial\, a^{(m)}}_{\theta} \leq \frac 1{\sqrt 2}$, showing by  ~(\ref{eq:3bis}) that $a^{(m)}$ is in the unit ball.  
 \end{proof}

Notice that the distance between nearest points
$\omega_{k-1}$ and $\omega_k$ is $\sqrt{\frac{\theta}{2k}}$, and the distance
between $\omega_p$ and $\omega_q$ ($p<q$) is the sum over the path joining the
two points
$$
d(\omega_p,\omega_q)=\sum\nolimits_{k=p+1}^qd(\omega_{k-1},\omega_k) \;.
$$
In other words, for any $p<n<q$, $\omega_n$ is a middle point between 
$\omega_p$ and $\omega_q$, i.e.~the triangle inequality becomes an equality
$d(\omega_p,\omega_q)=d(\omega_p,\omega_n)+d(\omega_n,\omega_q)$. 

Note also that the element $a^{(m)}\in\cal{S}(\gN)$ that attains the
supremum (defined in \eqref{eq:10}) has a geometrical interpretation, in analogy with the commutative case. 
 
\begin{Proposition} In the weak topology,
\begin{equation*}
  \lim_{m\rightarrow \infty } i[D,\pi(a^{(m)})] =  \begin{pmatrix}
0 & S^*\\ S& 0
\end{pmatrix} \;
\end{equation*}
where $
S=Z^*\abs{Z}^{-1},\,
S^*=\abs{Z}^{-1} Z
$ with $Z= L(z)$ the representation of $z$ 
on $\mathcal{H}_0$. 
\end{Proposition}
\begin{proof}

First notice that in
the matrix basis
$$
Zf_{m+1,n}=\sqrt{\theta (m+1)}f_{mn} \;,\qquad
Z^*f_{mn}=\sqrt{\theta(m+1)}f_{m+1,n} \;,
$$
for all $m,n\geq 0$, and $Zf_{0,n}=0$. Hence $\abs{Z} = (ZZ^*)^{\frac 12}$ acts as
\begin{equation*}
  \abs{Z} f_{mn}  = \sqrt{\theta(m+1)}f_{mn}
\end{equation*}
and is thus invertible, so that bounded operators $S, S^*$
are well defined.
Their action is $Sf_{m,n}=f_{m+1,n}$ and $S^*f_{m+1,n}=f_{m,n}$.

From (\ref{eq:12}) one has 
\begin{equation*}
(\partial\, a^{(l)}) f_{mn} = \sum_{p} \alpha_{p+1,p} \,f_{p+1,p} \star f_{mn} = \alpha_{m+1,m} \, f_{m+1,n} = \left\{
  \begin{array}{cc}
    \frac 1{\sqrt 2} \; S\,f_{mn} & \text{ for } m\leq l,\\
0& \text{ for } m > l,
  \end{array}\right.
\end{equation*}
hence the result by (\ref{eq:2}). 
\end{proof}
\noindent In the commutative limit, i.e.~$\theta=0$, $S=|z|^{-1}z^*$
is (almost everywhere) equal to $2\partial|z|=\sqrt
  2[\partial ,\sqrt 2\abs{z}]$.
The function $z\mapsto \sqrt 2\abs{z}$ is precisely the function that attains the supremum in the computation of the spectral distance on the positive
half real line, i.e.~$d(x,y)$ for $0\leq x\leq y$ (the $\sqrt 2$
factor comes from our definition (\ref{defz}) of $z$). In this sense
${\cal{P}(\cal{S}(\gN))}$  is a kind of discretization of the positive
half real line within the Moyal plane. See \cite{mm}.

\subsection{Lower bound and states at infinite distance}

In the precedent section we computed the distance between pure
  states associated to vectors \eqref{eq:39} with only one non-zero
  component. Studying vectors with more components is not easily
  tractable, due to the lack of an explicit formula for $\norm{[D,a]}$
  (instead, Proposition \ref{lemme2} only gives necessary conditions satisfied by the elements of the unit
  ball). Nevertheless we show in this section how to provide a lower
  bound for the distance between any two pure states, allowing to exhibit some states that are at infinite distance from one another.

We first notice that the difference between any two vector states of Proposition \ref{purestates} is naturally related to universal differential $d_Ua=a\otimes
\bbbone-\bbbone\otimes a, \, a\in{\cal{A}}$.
\begin{Lemma}
Let $\omega_{\psi^\prime}$ and $\omega_\psi$ be two pure state. For any $a\in{\cal{A}}$, one has
$$
\omega_{\psi^\prime}(a)-\omega_{\psi}(a)= (\omega_{\psi^\prime}\otimes \omega_{\psi} ) d_Ua = (2\pi\theta)^2\sum_{m,n,p,q}(d_Ua)_{mn,pq}\psi^{\prime *}_m\psi^*_p\psi^\prime_n\psi_q\ 
$$
where $(d_Ua)_{mn,pq}:=(a_{mn}\delta_{pq}-a_{pq}\delta_{mn}) $.
\end{Lemma}
\begin{proof}
One has
\begin{align*}
\omega_{\psi^\prime}(a)-\omega_{\psi}(a) &=2\pi\theta\sum_{m,n}a_{mn}(\psi^{\prime *}_m\psi^\prime_n-\psi^{ *}_m\psi_n)\\
&=(2\pi\theta)^2\sum_{m,n,p,q} (a_{mn}\delta_{pq}-a_{pq}\delta_{mn})\psi^{\prime *}_m\psi^*_p\psi^\prime_n\psi_q
\end{align*}
where the second equality stems from $||\psi^\prime||_2=||\psi||_2=1$. 
\end{proof}
\begin{Lemma}
For any unit vector $\psi,\psi'\in\hh_0$,  
\begin{equation}\label{eq:6}
d(\omega_\psi,\omega_{\psi'})\geq
(2\pi\theta)^2\sqrt{\frac{\theta}{2}}\; \Bigg|\sum_{q=1}^{\infty} \,\sum_{p=0}^{q-1} \,\sum_{k=p+1}^q\frac{1}{\sqrt{k}}
(|\psi_p\psi_q'|^2-|\psi_q\psi_p'|^2)\Bigg| \;
\end{equation}
whenever the r.h.s. is absolutely convergent or diverges to infinity. \end{Lemma}

\begin{proof}
Take $a^{(m)}\in \ub$ defined in \eqref{eq:10}. Since $a^{(m)}$ saturates (\ref{eq:16}), one has
\begin{align*}
\omega_\psi (a^{(m)})-\omega_{\psi'}(a^{(m)})
&=(2\pi\theta)^2\sum_{p,q}(a^{(m)}_{pp}-a^{(m)}_{qq})|\psi_p\psi_q'|^2 \\
&=(2\pi\theta)^2\sum_{p<q}(a^{(m)}_{pp}-a^{(m)}_{qq})(|\psi_p\psi_q'|^2-|\psi_q\psi_p'|^2) \\
&=(2\pi\theta)^2\sqrt{\frac{\theta}{2}}\sum_{p<k\leq\min(m+1,q)}\frac{1}{\sqrt{k}}
(|\psi_p\psi_q'|^2-|\psi_q\psi_p'|^2). \;
\end{align*}
The result then follows from $d(\omega_\psi,\omega_{\psi'})\geq
\lim_{m\rightarrow +\infty} \;\omega_\psi
(a^{(m)})-\omega_{\psi'}(a^{(m)})$,
whenever the limit exists (finite or not). 
\end{proof}

\begin{Proposition}
\label{propinfinie}
Consider the  two unit vectors $\psi, \psi'\in\hh_0$ with components
\begin{equation}\label{eq:7}
\psi_q=\frac 1{\sqrt{2\pi\theta}}\delta_{q0}\;,\quad\psi'_q=\frac 1{\sqrt{2\pi\theta}\sqrt{\zeta(s)q^s}} \quad\text{ for } q\neq 0,\, \psi'_0=0\;,
\end{equation}
where $s>1$ and $\zeta(s)$ is Riemann zeta function. If $s\leq 3/2$,
then $d(\omega_\psi,\omega_{\psi '})=\infty$.
\end{Proposition}

\begin{proof}
From \eqref{eq:6} we get
\begin{equation}
d(\omega_\psi,\omega_{\psi'})\geq
\sqrt{\frac{\theta}{2}}\frac{1}{\zeta(s)}\,\sum_{1\leq k\leq q}
\frac{1}{q^s\sqrt{k}}\geq
\sqrt{\frac{\theta}{2}}\frac{1}{\zeta(s)}\,\sum_{q=1}^\infty q^{\frac{1}{2} -s}
\label{eq:66}
\end{equation}
where we used $\sum_{k=1}^q \frac{1}{\sqrt{k}}\geq \sqrt{q}$. The r.h.s term of
\eqref{eq:66} diverges
for $s\leq 3/2$.
\end{proof}

Proposition \ref{propinfinie} shows that the diameter 
$d_\A =\text{sup} \left\{ d(x,y) ;\, x,y \in
  \mathcal{P}(\A)\right\}\label{eq:64}$
of the metric space $(\mathcal{P}(\A), d)$ is infinite, as the
diameter of the Euclidean plane. However the Euclidean distance between any two points in the plane is
always finite, although one may choose the points so that to make it 
arbitrarily large. The situation is different in the Moyal
plane: the diameter is infinite and the distance can also be infinite.  In this sense, the non-locality
of the Moyal product makes the distance larger.

So far, the only known cases where
the spectral distance between two states  $\omega,\omega'$  was
infinite were due to algebraic properties, namely the existence of a
non-trivial element $a_0$ such that
\begin{equation}
  \label{eq:58}
  [D,\pi(a_0)] = 0, \quad \omega(a_0) - \omega'(a_0) \neq 0.
\end{equation}
\noindent Considering $na_0$, $n\rightarrow +\infty$, one had that $d(\omega, \omega') = +\infty$. Here only constant functions commute with $D$ and the infinity of the distance has analytical origin, owing to the infinite dimension of the algebra.

\subsection{Spectral metric space}
In \cite{Rie99,Rie03,Rie04} Rieffel introduced
the notion of compact quantum metric spaces (that we recall below, cf.~Def.~\ref{qmsdef}) which  has been adapted to the non-compact case (i.e.~for non unital
algebras) by Latr\'emoli\`ere \cite{Lat05}. This leads to the recent
definition of \emph{spectral metric space}, namely quoting \cite{Bellissard:2010fk}: 
${\small \ll}\,$A spectral metric space is a spectral triple $(\A,\HH,D)$ with
additional properties which guaranty that the Connes
metric{\footnote{That here we call spectral distance.}} induces the
weak*-topology on the state space of $\A\,{\small \gg}$. 

When the unit ball (\ref{unitball}) is bounded these properties have
been established in \cite{Lat05}, leading to the definition of
\emph{bounded} spectral metric space in \cite{Bellissard:2010fk}. When
the unit ball is not bounded, the question is still open. In
\cite[Prop.5]{Bellissard:2010fk} an example of
spectral triples based on $\Z$ is introduced that
fails to be a spectral metric space as soon as the unit ball
is not norm bounded. A similar
situation occurs for the Moyal plane: the unit ball is not norm bounded
since $\norm{a^{(m)}}_{\text{op}}\to\infty$ (see (\ref{eq:10})) and we
show below that the topology induced by the spectral distance is not
the weak* one.

\begin{Lemma}\label{lemma-pd}
Let ${\mathcal X}$ be a topological space whose topology is induced by a pseudo-distance $d$~{\footnote{Following \cite{Villani:2009tp},
a pseudo-distance is a function that satisfies all the properties of a distance, except that it may be infinite.}}. 
Then, if $d(\varphi_0, \varphi_1)=\infty$ the points $\varphi_0, \varphi_1\in\mathcal{X}$ are in 
distinct connected components.
\end{Lemma}
\begin{proof}
Let us fix $\varphi_0\in{\mathcal X}$. Call  
\begin{equation*}
S_0 \doteq \{\varphi\in {\mathcal X},\;
d(\varphi_0, \varphi)<\infty\},
\end{equation*}
and $B_r(\varphi)\doteq \left\{\varphi' \in {\mathcal X}, d(\varphi, \varphi')<
r\right\}$ the open ball or radius $r>0$ centered at $\varphi$. For any
$\varphi\in S_0$ and $r>0$, one has $B_r(\varphi)\subset S_0$. Indeed if
$\varphi'\in B_r(\varphi)$, by the
triangle inequality 
\begin{equation*}
  d(\varphi_0, \varphi')\leq d(\varphi_0, \varphi) + d(\varphi,
  \varphi') <\infty,
\end{equation*}
  so that $\varphi'\in S_0$. Hence $S_0$ is open.
Similarly, the inequality
\begin{equation*}
  d(\varphi_0, \varphi)\geq d(\varphi_0, \varphi') - d(\varphi,
  \varphi')= \infty
\end{equation*}
for all $\varphi'\notin S_0$ and $\varphi\in B_r(\varphi')$, shows
that the complement of $S_0$ is open too. Hence $S_0$ is closed.
Any clopen set is a union of connected components, and this proves
that $\varphi_0\in S_0$ and $\varphi_1\notin S_0$ cannot be in the same
connected component.
\end{proof}

Given two states $\varphi_0, \varphi_1\in \sa$, consider the map
\begin{equation}
[0,1]\ni
t\mapsto \varphi_t = (1-t)\varphi_0 +  t\varphi_1\in \sa.
\label{eq:8}
\end{equation}
The spectral distance satisfies \cite[eq.~(1.9)]{DM09}
\begin{equation*}
    d_D (\varphi_t , \varphi_{t'}) = |t-t'| d_D (\varphi_0 , \varphi_1),
    \quad \forall\,  0 \leq t,t'\leq 1.
\end{equation*}
This implies that the map in \eqref{eq:8} is continuous for the
topology induced by the spectral distance if $d_D(\varphi_0,\varphi_1)<\infty$. Therefore,

\begin{Lemma}\label{lemma:3.12}
If $d_D(\varphi_0,\varphi_1)<\infty$, then $\varphi_0, \varphi_1\in \sa$ are in the same
connected components for the topology induced by the spectral distance. Moreover, connected
components are path connected.
\end{Lemma}

\begin{Lemma}\label{lemma:3.13}
$\sa$ is path-connected for the weak* topology.
\end{Lemma}
\begin{proof}
Recall that a \emph{density matrix} $P$ (on $\ell^2(\N)$) is a positive trace-class operators on $\ell^2(\N)$ with trace $1$. A \emph{normal} state $\omega_P\in\sa$ is a state that
can be written as
$$
\omega_P(a)=\tr(Pa)\;.
$$
Since $\bar{\A}\simeq\mathbb{K}$, it follows from Prop.~2.6.13
of \cite{BR97} that any state of the Moyal algebra is normal, that is 
$\sa$ is identified with the set of density matrices.
By \cite [Prop.~2.6.15]{BR97}, the weak* topology on $\sa$ is
equivalent to the uniform topology induced by the trace norm,
\begin{equation}
  \label{eq:1}
  \norm{T}_{\text{Tr}} \doteq \tr\, \abs{T} \quad\text{ for all traceclass } T.
\end{equation}
Similarly to \eqref{eq:8}, for any density matrices $P_0,P_1$
we define a map
$$
[0,1]\ni t\mapsto P_t = (1-t)P_0 +  tP_1 \;.
$$
For all $0\leq t,t'\leq 1$ we have $P_t-P_{t'} =(t-t')(P_1-P_0)$ so that
\begin{equation}
\label{eq:geoHS} 
\norm{P_t - P_{t'}}_{\text{Tr}} =|t-t'|\norm{P_0 - P_1}_{\text{Tr}} \;.
\end{equation}
Therefore,  $t\mapsto P_t$ is continuous in the weak* topology
(for all $\epsilon>0$ called $\delta_\epsilon=\epsilon/\norm{P_0 - P_1}_{\text{Tr}}$ from \eqref{eq:geoHS} we get
$|t-t'|<\delta_\epsilon \Rightarrow \norm{P_t - P_{t'}}_{\text{Tr}} <\epsilon$) and so $\sa$
is path-connected, as claimed.
\end{proof}

\begin{Proposition}\label{spectmet}
 The topology induced by the spectral distance is not
the weak* topology, thus the spectral triple introduced in \cite{Gayral:2004rc} for the Moyal
  plane is not a spectral metric space. 
\end{Proposition}
\begin{proof}
Suppose that two topologies $T_1$ and $T_2$ are equivalent:
then, if $S$ is a connected component of $T_1$, it must be also a
connected component of $T_2$. 
Let $T_1$ be the topology induced on $\sa$ by the spectral distance
and $T_2$ the weak* topology. By Lemma \ref{lemma-pd} the two pure states in
Prop.~\ref{propinfinie} are in different connected components of $T_1$.
On the other hand, from Lemma \ref{lemma:3.13} it follows that 
any two pure states are in the same connected component for the weak$^*$ topology
($\sa$ is path-connected). This concludes the proof.
\end{proof}

\section{Truncated Moyal space as compact quantum metric spaces}
\label{sec:4}

\subsection{Quantum metric spaces}
In \cite{Rie03} Rieffel introduces the notion of quantum metric space motivated on the one side by Connes distance formula \eqref{spectralformule}, on the other side by the the observation that on a compact Hausdorff metric space  $(X,\rho)$ with Lipschitz seminorm
$l(f)=\sup_{x\neq y}|f(x)-f(y)|/\rho(x,y)\,$
  on $A=C(X)$ (the infinite value is permitted), one can define on ${\cal S}(A)$ a distance 
\begin{equation}
  \label{eq:21}
  \rho_l(\omega_1, \omega_2)  := \sup\left\{\omega_1(f)- \omega_2(f); \, l(f)\leq 1\right\}
\end{equation}
whose topology  coincides with the weak* topology,
\begin{equation*}
  \lim_{n\rightarrow \infty}\rho_l (\omega_n -\omega) = 0 \;\,\text{ iff }\;\,  \lim_{n\rightarrow \infty}\,\omega_n(f) -\omega(f) = 0 \quad \forall f\in A,
\end{equation*}
and whose restriction
on ${\cal P}(A)$ gives back $\rho$. When moreover $X$ is a Riemannian spin manifold with Dirac operator $D$, then
$l(f) = \norm{[D,f]}$ and \eqref{eq:21} is nothing but the spectral distance. Say differently, for a compact commutative spectral triple the topology defined by the spectral distance on the state space coincides with the weak* topology. However there is a priori no reason that this still holds true for arbitrary spectral triples. This motivates the following definition of \emph{compact quantum metric space}.

\begin{Definition}\label{qmsdef}\cite{Rie03}
A {\it compact quantum metric space} is an order-unit
  space $A$ equipped with a seminorm $l$ such that 
$ l(1)=0$ and the distance
\begin{equation}
\label{eq:46} 
d(\omega_1, \omega_2):=\sup\big\{ \omega_1 (a)-\omega_2 (a) \,;\, l(a)\leq 1 \big\}
\end{equation}
induces the weak* topology on the state space of $A$.
\end{Definition}
\noindent Note that for technical flexibility Rieffel \cite{Rie99} uses order-unit spaces rather than algebras. The precise definition can be found in \cite{Alf71, Kad51}. For our purpose
we just need to know that any real linear space of self-adjoint operators on a Hilbert space
containing the identity operator $\I$ is an order-unit space, and
any order-unit space can be realized in this way. Moreover the notion of
state naturally extends to order-unit spaces. Therefore, since  the supremum in (\ref{spectralformule}) can be searched on selfadjoint elements, it make sense to view a unital spectral triple whose spectral distance induces on ${\cal S}(A)$  the weak$^*$ topology as a  compact quantum metric space, with seminorm $l(.) = \norm{[D,.]}.$

A necessary condition \cite{Rie03}  for a semi-norm $l$ on $A$
  to define a quantum metric space is
\begin{equation}
l(a) = 0 \Longleftrightarrow a\in \gR \I.
\label{eq:45}
\end{equation}
Indeed when (\ref{eq:45}) does not hold,  it is not difficult to 
find states at infinite distance from one another (see \eqref{eq:58}) so that
the metric and weak$^*$ topologies cannot coincide since ${\mathcal
  S}(A)$ --- for any order-unit space $A$ --- is compact for the weak$^*$
topology. However this condition is not sufficient: consider the
order-unit space $\A_1^{sa}$ of the Moyal algebra introduced in section
\ref{spectraltriple} with seminorm $\norm{[D,.]}$ and associated
distance $d_1$. The unit vectors
$\psi, \psi'$ in  (\ref{eq:7}) still define pure states $\omega_\psi,
\omega_{\psi'}$ of $\A_1$ and,
since $\A\subset\A_1$,
one has $d\leq d_1$. In particular by Proposition \ref{propinfinie}
$d_1( \omega_\psi,\omega_{\psi'})= +\infty$, although (\ref{eq:45})
holds true. 
For this reason the Moyal plane equipped with the distance $d_1$ is
\emph{not} a quantum metric space.

However for spectral triples whose algebra has finite
  dimension, (\ref{eq:45}) guarantees that the spectral distance
  induces the
weak$^*$ topology. Let us recall  that any finite dimensional
$C^*$-algebra $\A$ is a direct sum of matrix
 algebras,
$\A = \sum_{n\in I} M_n(\C)$,
and $\pa =\cup_{n\in I}{\mathcal P}(\A_n)$ where ${\mathcal P}(\A_n)$ is the
set of vector states
\begin{equation}
a\in M_n(\C) \mapsto \omega_\Psi(a) = \langle \Psi, a\Psi \rangle= \text{Tr} (s_\Psi a)\label{eq:71}
  \end{equation}
where $\Psi$ is a unit vector in $\C^{n}$, $s_\Psi =
|\Psi\rangle\langle \Psi|\in M_n(\C)$ is the associated
projection and the omission of the symbol $\pi$ means that one is
considering the fundamental representation. $\caS(\A)$  is in $1$-to-$1$ correspondence with convex
sums of rank $1$ projections on each $M_n(\C)$, i.e.
$$
\caS(\A)= \left\{  s\in \A^+,\, \text{Tr}\, s = 1\right\} : \A\ni a\mapsto \text{Tr}(sa)
$$
where $\A^+$ denotes the set of positive elements of $\A$. In physicist
language, these are the density matrices and one recovers that a state
is pure iff the associated density matrix is a projection in a single
component $M_n(\C)$.

\begin{Proposition}\label{quantumetrictheo}
A spectral triple $(\A, \hh, D)$, with $\A$ a finite dimensional algebra, is a compact quantum metric space
iff
\begin{equation}
[D,\pi(a)]= 0  \Longleftrightarrow a = \lambda \mathbb{I} \quad \text{
  for } \lambda\in\C.\label{eq:67}
\end{equation}
 \end{Proposition}
\begin{proof}
 i) Assume \eqref{eq:67} does not hold. Consider $b_0\neq \lambda
 \mathbb{I}$ such that $[D,\pi(b_0)]=0$. For $a_0 =\frac 12 (b_0 + b_0^*)\in \mathcal{B}_D^{sa}$,
$\pi(a_0)$ has at least two normalised eigenvectors $\psi,\psi'\in\hh$ with distinct
eigenvalues $\alpha, \alpha '$. By \eqref{eq:58} the corresponding
states are at infinite distance
since $\omega_{\Psi}(a_0) = \alpha\neq \alpha'=
\omega_{\Psi'}(a_0)$. Hence the spectral triple is not a
compact quantum metric space.

ii) ii) Assume that \eqref{eq:67} holds. Two distances $d_1$ and $d_2$  induce the same
topology on $\caS(\A)$  if they are strongly equivalent,  that is to say  if there exist
two constants $C,C'>0$ such that
\begin{equation}
C\,d_1(r,s)\leq d_2(r,s)\leq C'\,d_1(r,s) \quad \forall r,s \in \caS(\A).
\label{eq:49}
\end{equation}
In particular two distances defined via (\ref{eq:46}) by 
semi-norms $l_1, l_2$ satisfying 
\begin{equation}
K'\, l_1(a)\leq l_2( a)\leq K\, l_1(a) \quad \forall a\in\A \label{eq:29}
\end{equation}
 for some constant $K, K'$ are equivalent with $C' = K'^{-1}, C =
 K^{-1}.$ 
By \eqref{eq:3bis}, Remark \ref{autoadjoint} and noting that 
adding a multiple
of $\mathbb{I}$ to $a$ doesn't change $\tr(ra) -\tr(sa)$ nor $[D,\pi(a)]$, 
one has
\begin{equation*}%\label{eq:3}
d(r,s)=\sup_{ a\in\V\cap {\mathcal B}_D}\big\{\tr(ra)-\tr(sa)\big\} \;
\end{equation*}
where $\V$ denotes the set of selfadjoint $a\in\A$ 
such that $\tr(a)=0$. 
Moreover $||[D,\pi(.)]||$  on the vector space $\V$ is a
norm since 
\begin{equation*}
||[D,\pi(a)]||_{\mathrm{op}}=0 \Longrightarrow  
a=\lambda \mathbb{I}
\end{equation*}
by \eqref{eq:67} and $\lambda=n^{-1}\tr(a)=0$ with $n$ the
  dimension of $\A$. All norms on
a finite dimensional vector space being equivalent, the first part
of the proposition follows from \eqref{eq:49} and \eqref{eq:29} as soon as one exhibits a norm on $\V$ whose
associated distance induces the weak$^*$ topology. 

The $L^2$ norm  $||a||_{L_2} := \sqrt{\inner{a,a}}$ is
such a norm, where $\inner{a,b} = \tr(a^*b)$. Indeed, defining 
\begin{equation*}
d_2(r,s)\doteq \sup_{a\in\V}\big\{ \tr(ra)-\tr(sa) ,\;||a||_{L_2}\leq 1 \big\},
\end{equation*}
one has that
\begin{equation}\label{eq:280}
d_2(r,s)=||r-s||_{L_2}
\end{equation}
since, by Cauchy-Schwarz and for any $a\in\V$ such that $||a||_{L_2}\leq 1$, 
$$
\tr(ra)-\tr(sa)=\inner{r-s,a}
\leq 
||a||_{L_2}||r-s||_{L_2}
\leq ||r-s||_{L_2}
$$
and the equality is attained by
the unit-norm element $||r-s||_{L_2}^{-1}(r-s)\in  \V$.
(\ref{eq:280}) obviously induces the weak$^*$ topology: 
a sequence of density matrices $s_n$  with components $(s_n)_{ij}$
tends to $s$, i.e.~$\lim_{n\rightarrow +\infty}||s_n-s||_{L_2} = 0$,
iff 
$
|(s_n)_{ij}-s_{ij}|\to 0
$ for all $i,j$.  
\end{proof}

Notice that Prop.~\ref{quantumetrictheo} can also be derived from
Theorem 4.5 of \cite{Rie04}, but the direct proof given
here may be of interest. 
% %%% -----------------------------------------------------------------------
\subsection{Truncation of the Moyal spectral triple}\label{sectionweak}

Viewing $\A$ as the inductive limit 
$$
\A  = \overline{\underset{\longrightarrow}{\lim }\,\A_n}, \quad \A_n :=M_n(\C)
$$
with morphism the natural embedding of $M_n(\C)$ into $M_{n+1}(\C)$
and here the closure is with respect the Fr\'echet norm (\ref{eq:57}), we
show below how the restriction of the spectral triple of the Moyal
plane to
a finite rank $n$ yields a compact quantum metric space. The
truncation of the spectral triple (\ref{eq:13})  is the following: the algebra is $\A _n$, acting on $\hh _n = M_n(\gC) \otimes \gC ^2$
as $L(a)\otimes{\mathbb I} _2$ where $L(a)$ is the left regular representation of $M_n(\gC)$ on itself, 
with inner product normalized as
$
\inner{a,b}_{\theta}=2\pi\theta\,\mathrm{Tr}(a^\dag b) \;.
$
The Dirac operator (\ref{eq:14}) is now defined by the two derivations
\begin{equation}
\partial a=-[X_-,a] \;,\qquad
\bar\partial a=[X_+,a] \;,
\label{eq:61}
\end{equation}
with
\begin{equation*}
X_-:=\frac{1}{\sqrt{\theta}}\begin{pmatrix}
0 & 0 & 0 & \ldots&0 \\
1 & 0 & 0 & \ldots &0\\
0 &\sqrt{2} & 0 & \ldots&0 \\
\vdots & \ddots & \ddots & \ddots &\vdots\\
0 & \ldots& 0 & \sqrt{n-1}& 0
\end{pmatrix}
\;,\qquad
X_+:=X_-^t .
\end{equation*}
\noindent 
For $n=\infty$ one recovers the Dirac operator of the Moyal plane.

With notations of Proposition \ref{purestates}, ${\cal P}(\A_n)$ is the set of vector states
\begin{equation*}
\omega_\psi(a) = \langle \psi, a\psi\rangle_\theta = \langle s_\psi ,a \rangle_\theta
\end{equation*}
with
\begin{equation*}
\psi = \left(\begin{array}{cccc} \psi_1 & 0 & \ldots & 0 \\
    \psi_2&0 & \ldots& 0\\ 
\vdots&\vdots & \vdots& \vdots\\
\psi_n&0 & 0&  0\\\end{array}\right), \; s_\psi = \psi\psi^*
\end{equation*}
where $\psi_i$ are complex numbers such that $ \langle \psi,
\psi\rangle_\theta = 2\pi\theta\sum_{j=1}^{n} \abs{\psi_j}^2 =1$.  Two matrices $\psi, \psi'\in M_n(\C)$ define the same
  pure state iff $\psi\psi^* = \psi'\psi'^*$,  e.g.~--- with notation of
  (\ref{eq:39}) --- $\psi = f_{mp},
\psi'=f_{mq}$. Writing
$\Psi\in\C^2$ the unit vector with component $\sqrt{2\pi\theta}\, \psi_i$, one retrieves the usual
form \eqref{eq:71} with  $\langle
., . \rangle$ the inner product on $\gC^n$ and $s_\Psi = 2\pi\theta s_\psi$ the
projection on $\Psi$. Two unit vectors equal up to a phase define the
same state, hence ${\cal P}(\A_n) = S^{2n-1}/S^1 = \gC{\mathbb
P}^{n-1}.$ Note that the ambiguity in the choice of $\psi$ is $U(n)$,
while the ambiguity in the choice of $\Psi$ is $U(1)$. 

\begin{Proposition}
\label{truncprop}
The truncated Moyal plane $(\A_n, \hh_n, D)$ is a compact quantum
metric space. With the distance $d_2$
introduced in (\ref{eq:280}), $\caS(\A_n)$ has radius smaller than $2\sqrt{1-\frac
  1n}$. 
\end{Proposition}
\begin{proof}
The first statement follows from Proposition \ref{quantumetrictheo}, noting that nothing
but multiples of the identity commute with $D$. 
For any state $s\in{\mathcal S}(\A)$,
\begin{equation}
d_2(\tfrac 1n {\mathbb I},s)^2=\tr((\tfrac 1n {\mathbb I}-
  s)^2)= \tr s^2 -\tfrac 1n \leq 1-   \tfrac 1n
\label{eq:72}
\end{equation}
where we noticed that for any positive matrix $s$ of trace $1$,
$\text{Tr } s^2 \leq \text{Tr } s$. Hence
\begin{equation*}
d_2(r,s) \leq d_2(r,\tfrac 1n\mathbb{I}) +d_2(\tfrac
1n\mathbb{I},s) \leq 2\sqrt{1 -\tfrac 1n}
\end{equation*}
for all $r,s\in \mathcal{S}(\A_n).$\end{proof}

The upper bound in (\ref{eq:72}) is attained by projections
{\footnote{Let $s$ be such that
$\text{Tr } s^2 = 1 = \text{Tr } s$.  Writing $0\leq \alpha_i\leq 1$
the eigenvalues of $s$, $\text{Tr } s^2 = \text{Tr } s$ yields
$ \sum_{i=1}^N \alpha_i-\alpha_i^2 = 0$. Each term of the sum being
positive, the sum  is zero
 iff $\alpha_i = \alpha_i^2$ for any $i$, i.e $\alpha_i = 0$ or
 $1$. Since the trace is $1$, this means all $\alpha_i$ vanish except
 one. Hence $s$ is a projection.}}, so that ${\cal P}(\A_n)$ is the
sphere of radius $\sqrt{1 -\frac 1n}$ centered on $\frac
1n\mathbb{I}$, and $\caS(\A_n)$ is the corresponding
ball. Alternatively, one may extend the distance $d_2$ 
to $\A^+_n$, yielding a characterization of the state space in term of the unit sphere $S^1$
 and the unit ball $B^1$ of the metric space $(\A^+_n,
 d_2)$. With calculations similar to the ones in Proposition
 \ref{truncprop}, one gets that 
%\begin{itemize}
%\item 
${\cal P}(\A_n)$ is the intersection of $S^1$  with the set of
matrices of trace $1$; and
 %\item 
${\cal S}(\A_n)$ is the intersection of $B^1$ with the set of matrices of trace $1$.
%\end{itemize}

With some more calculation, one finds $C d_2 \leq d \leq C' d_2$
with 
\begin{equation}
C\leq \tfrac{1}{2}\sqrt{\tfrac{\theta}{2n}},\quad C'\geq (2n^3+n^2)\sqrt{\tfrac{\theta}{2}}.\label{eq:36}
\end{equation}

Proposition \ref{truncprop} is particularly interesting in
comparison with other spectral distances on $M_n(\C)$:
  in \cite{IKM01} one takes as a Dirac operator $D(a) := M a + aM$ with $M=M^*\in
  M_n(\gC)$. This choice is physically relevant since the
  coefficients of $M$ can be interpreted as masses, but the spectral triple does not yield a compact
  quantum
  metric space: condition (\ref{eq:45}) is not satisfied --- the
  eigenprojections of $M$ commute with $D$ ---  and ${\cal P}(\A_n)$
  decomposes into sub-tori, all at infinite distance from one
  another. In \cite{Christensen:2006fk} one gets a compact metric space 
using the operator of matrix transposition as Dirac
  operator (in this case the spectral distance coincides with the
  operator norm); 
but there is no room in the Dirac operator to put mass
  parameters ($D(a)={^t}(a)$ has only eigenvalues $\pm 1$).
With Proposition \ref{truncprop} one combines $n$ arbitrary
masses (the proposition still holds with arbitrarily non-zero
values instead of $1,\ldots,\sqrt{n-1}$ in $X_-$) with a metric
giving the weak$*$ topology on $\sa$.

\subsection{A  spectral distance on the $2$-sphere}

To illustrate our
  results, we close this paper by explicitly computing the distance
  associated to the truncated spectral triple for $n=2$. The algebra $\A_2 = M_2(\gC)$ acts on $M_2(\gC)\otimes \gC^2$
as block diagonal matrices and $D$ is given by \eqref{eq:61} with
\begin{equation*}
  \partial a = -\frac 1{\sqrt \theta}\left[\left(\begin{array}{cc} 0&0\\ 1&0 \end{array}\right),a\right], \quad  \bar\partial a = (\partial a)^*\quad\quad \forall a=a^*\in \A_2.
\end{equation*}
The pure states space ${\cal P}(\A_2)= \C\mathbb{P}^1$ is identified to the sphere $S^2$
via the map
\begin{equation*}
\Psi =\binom{\Psi_1 }{ \Psi_2} \quad\longleftrightarrow\quad
\left\{\begin{array}{l}
x_\Psi :=2\,\text{Re} \bar\Psi_1 \Psi_2 \\[2pt]
y_\Psi :=2\,\text{Im} \bar\Psi_1 \Psi_2 \\[2pt]
z_\Psi :=\abs{\Psi_1}^2 -\abs{\Psi_1}^2.
\end{array}\right.
\end{equation*}
The evaluation on $a\in\A$ with components $a_{ij}$ reads
\begin{equation*}
  \op(a) = \langle \Psi, a\Psi\rangle = \frac{1+\zp }2 a_{00} + \frac{1- \zp }2 a_{11} + r\,\Re \left(e^{i\Xi} a_{01}\right)
\end{equation*}
 with $r e^{i\Xi} := x_\Psi + i y_\Psi$.
A non-pure state $\omega_\phi$ is given by a probability distribution $\phi$ on $S^2$,
\begin{equation}
  \label{eq:38}
  \omega_\phi (a) = \int_{S^2} \phi(\xi) \omega_\xi(a) d\xi
 = \frac{1+\tilde z_\phi }2 a_{00} + \frac{1- \tilde z_\phi }2 a_{11} + \tilde r_\phi\,\Re \left(e^{i\tilde\Xi_\phi} a_{01}\right)
\end{equation}
where 
$d\xi$ is the $SU(2)$ invariant measure on $S^2$ normalized to $1$ and ${\bf \tilde x}_\phi = (\tilde x_\phi,  \tilde y_\phi,  \tilde z_\phi)$ denotes the mean point of $\phi$, with
\begin{equation*}
  \tilde x_\phi  := \int_{S^2} \phi(\xi) x_\xi d\xi
\end{equation*}
and similar notation for $\tilde y_\phi$, $\tilde z_\phi$.
The point ${\bf \tilde x_\phi}$ is in the unit ball $B^2$.  When $\of
=\op$ is  pure, then $\phi= \delta_{\xi - \psi}$ and one retrieves
${\bf \tilde x}_\phi = {\bf  x}_\Psi := (x_\Psi, y_\Psi, z_\Psi)\in
S^2$. The correspondences $\pad\leftrightarrow S^2$ and $\sad
\leftrightarrow B^2$ are 1-to-1. 

Noting that the 
weak$^*$-topology on $\caS(\A_n)$ coincides with the Euclidean topology on
${\mathcal B}^2$,
\begin{equation*}
  \lim_{n\rightarrow +\infty} \omega_{\phi^n} = \omega_{\phi} \;\;\Longleftrightarrow\;\; \lim_{n\rightarrow +\infty}\tilde{\mathbf{x}}_{\phi^n}=
 \tilde{\mathbf{x}}_\phi,
\end{equation*}
Proposition
\ref{truncprop} indicates that ${\cal S}(\A_2)$ with the topology induced by the spectral distance is homeomorphic to the Euclidean closed ball $B^2$.
${\cal P}(\A_2)$ is homeomorphic to the Euclidean sphere $S^2$.

Although they induce the same topologies, $d_2$, the Euclidean and the
spectral distances are not equal.
 Writing
\begin{equation*}
  s_\phi= \left( \begin{array}{cc}
\frac{z_\phi+1}2 & \frac{x_\phi-iy_\phi}2 \\ \frac{x_\phi+ iy_\phi}2 & \frac{1-z_\phi}2\end{array}\right)
\end{equation*}
the density matrix associated to $\omega_\Phi$, an easy calculation shows
that
\begin{equation}
  \label{eq:68}
  d_2(s_\phi, s_{\phi'}) =\frac 1{\sqrt 2} d_{Ec} ({\bf \tilde  x_\phi}, {\bf \tilde  x_{\phi'}}) 
\end{equation}
where
$$
 d_{Ec} ({\bf \tilde  x_\phi}, {\bf \tilde  x_{\phi'}}) =  \sqrt{\abs{\tilde x_\phi - \tilde x_{\phi'}}^2 + \abs{\tilde y_\phi - \tilde y_{\phi'}}^2+ \abs{\tilde z_\phi - \tilde z_{\phi'}}^2 }
$$
denotes the euclidean distance on $B^2$.  Note that the radius  $\frac
2{\sqrt 2}$ of
$\caS(\A_n)$ is exactly the upper bound in \eqref{eq:72}.
\begin{Proposition}
\label{propdista2} The spectral distance between any two states of ${\cal S}(\A_2)$ is
\begin{equation*}
d(\omega_\phi,\omega_{\phi'})=
\sqrt{\frac{\theta}{2}}\times\begin{cases}
d_{eq} ({\bf \tilde  x_\phi}, {\bf \tilde  x_{\phi'}}) & \mathrm{if}\;\;|\tilde z_\phi-\tilde z_{\phi'}|\leq d_{eq} ({\bf \tilde  x_\phi}, {\bf \tilde  x_{\phi'}}) \;,\\
\frac{d_{Ec} ({\bf \tilde  x_\phi}, {\bf \tilde  x_{\phi'}})^2}{2 |\tilde z_\phi-\tilde z_{\phi'}|}  & \mathrm{if}\;\;|\tilde z_\phi-\tilde z_{\phi'}|\geq d_{eq} ({\bf \tilde  x_\phi}, {\bf \tilde  x_{\phi'}}),  \;
\end{cases}
\end{equation*}
 where
$$
 d_{eq} ({\bf \tilde  x_\phi}, {\bf \tilde  x_{\phi'}})=  \sqrt{\abs{\tilde x_\phi - \tilde x_{\phi'}}^2 + \abs{\tilde y_\phi - \tilde y_{\phi'}}^2 }
$$
is the Euclidean distance between the projections of the points on the equatorial plane $z=0$.
\end{Proposition}

\begin{proof}
With notations of \eqref{eq:38} and by \eqref{eq:3bis},
\begin{equation}
d(\omega_\phi,\omega_{\phi'})=\sup_{a\in\V_2}\big\{
\of(a) - \ofp(a), \,
||\partial a||_{\mathrm{op}} = 2^{-\frac 12}
 \big\} \;.\label{eq:35}
\end{equation}
Any $a\in\V _2$ is of the form
\begin{equation*}
a=\begin{pmatrix}
B & A \\ \bar A & -B
\end{pmatrix}
\end{equation*}
with $A\in \gC$, $B\in\gR$, and has derivative
$$
\partial a=\frac{1}{\sqrt{\theta}}
\begin{pmatrix}
A & 0 \\ 
-2B & -A
\end{pmatrix} \;
$$
whose norm can  be explicitly computed as
  $\norm{\partial a}_\text{op} =\theta^{-\frac{1}{2}}\big(\sqrt{|A|^2+|B|^2}+|B| \big).$ 
Writing $\abs{B} + i\abs{A} = re^{i\lambda}$, one thus gets 
\begin{equation}
  \label{eq:30}
  ||\partial a||_{\mathrm{op}} = 2^{-\frac 12} \Longleftrightarrow r=\sqrt{\frac{\theta}2}\frac 1{1+\cos \lambda}. 
\end{equation} 
Noting by (\ref{eq:38}) that
\begin{equation*}
\of(a) - \ofp(a)=B Z+\Re(AX)\label{bzrx}
\;\text{ with }\; Z:= \tilde z_\phi-\tilde z_{\phi'},\,  X:= \tilde r' e^{i\tilde\Xi'}-\tilde r e^{i\tilde\Xi},
\end{equation*}
(\ref{eq:30}) together with \eqref{eq:35} yields
\begin{align*}
   d(\of,\ofp)& \leq\sup_{\abs{A}, \abs{B}\in\gR^+} \left\{ 
\abs{B}\abs{Z} + \abs{A}\abs{X},\;\sqrt{|A|^2+|B|^2}+|B| = \sqrt{\frac{\theta}2}\right\},\\ \nonumber
&\leq  \sup_{0\leq \lambda\leq \frac{\pi}2} \sqrt{\frac{\theta}2}\frac{\abs{Z}\cos \lambda+\abs{X}\sin \lambda}{1+\cos\lambda}\\ &
=  \sqrt{\frac{\theta}2}\times\left\{\begin{array}{cc}
\abs{X} &\text{ when } \abs{Z}\leq \abs{X}\\
 \frac{\abs{X}^2 + \abs{Z}^2}{2\abs{Z}} &\text{ when } \abs{Z}\geq \abs{X}.
\end{array}\right. 
\end{align*}
These upper bounds are attained by 
%\begin{align}&
$$B=0,\,  A = \sqrt{\frac{\theta}2} e^{-i\theta_X}, \,\theta_X := \text{Arg} X$$   in case $\abs{Z}\leq \abs{X}$ (which corresponds to $\lambda = \frac{\pi}2)$, and
$$B= (\text{sign }  Z)\frac{\abs{Z}^2 - \abs{X}^2}{2\abs{Z}^2}, \,A = \frac{\abs{X}}{\abs{Z}}$$
 when $\abs{Z}\geq \abs{X}$
(which corresponds to $\lambda = \arccos (\frac{\abs{Z}^2-\abs{X}^2}{\abs{Z}^2+\abs{X}^2})$.
The final result is obtained by noting that $\abs{X} =d_{eq} ({\bf \tilde  x_\phi}, {\bf \tilde  x_{\phi'}}) $.  
\end{proof}
\vspace{-0truecm}
\begin{figure}[ht*]
\begin{center}
\hspace{-0truecm}
\vspace{-1truecm}
\mbox{\rotatebox{0}{\scalebox{.8
}{\includegraphics{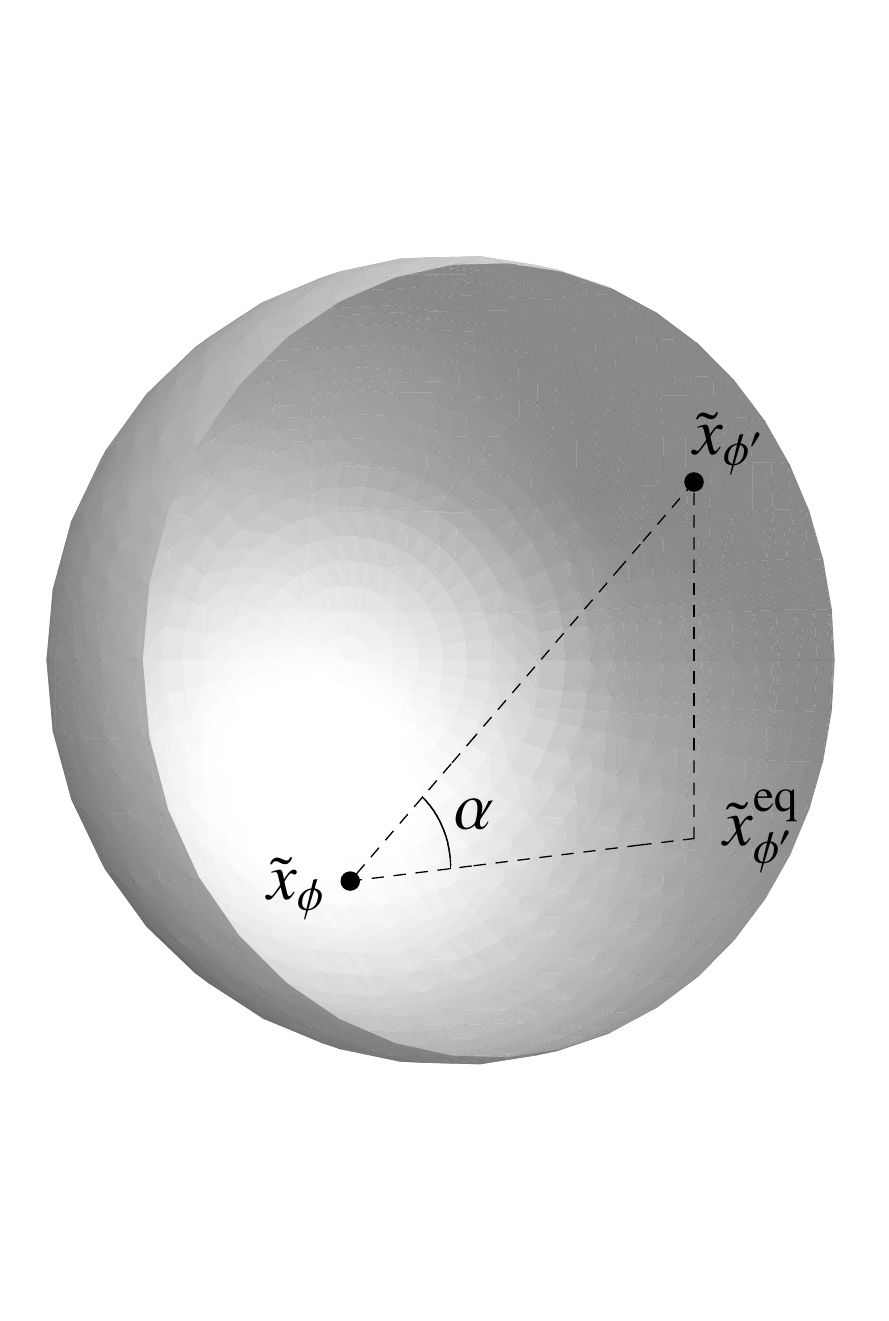}}}}
\end{center}
\caption{The vertical plane containing ${\bf \tilde  x_\phi}$, ${\bf \tilde  x_{\phi'}}$.}\label{Moyalfig}
\end{figure} 
\noindent This result has some analogy with the spectral distance computed on ${\cal P}(M_2(\gC))$ in \cite{IKM01}, namely the rotation around the $z$-axis is an isometry and the distance on a parallel ($\zp = \zpp$) is proportional to the Euclidean distance within the disk.  But remarkably,  while the distance on a meridian was infinite in \cite{IKM01}, it remains finite here. This allows an easy ``classical geometry'' interpretation
of Prop. \ref{propdista2}. 
Assuming for convenience that $\tilde z_\phi\geq \tilde z_{\phi'}$, let us call ${\bf \tilde  x_{\phi'}}^{eq}$ the projection 
of ${\bf \tilde  x_{\phi'}}$ in the $\tilde z_{\phi'}$ plane and $\alpha$ the angle $({\bf \tilde  x_{\phi}}{\bf \tilde  x_{\phi'}}^{eq}, {\bf \tilde  x_{\phi}}{\bf \tilde  x_{\phi'}})$ (see figure \ref{Moyalfig}).
Then 
\begin{equation*}
d(\omega_\phi,\omega_{\phi'})=
\sqrt{\frac{\theta}{2}}\times\begin{cases}
\cos\alpha\;\, d_{Ec} ( {\bf \tilde  x_\phi}, {\bf \tilde  x_{\phi'}}) & \mathrm{when }\;\,\alpha\leq \frac{\pi}4,\\
\frac{1}{2\sin \alpha}\, d_{Ec} ({\bf \tilde  x_\phi}, {\bf \tilde  x_{\phi'}})  & \mathrm{when }\;\, \alpha\geq \frac{\pi}4.
\end{cases}
\end{equation*}
Note that by \eqref{eq:68} this result is in agreement with
\eqref{eq:36}. Also,  two states with same $\tilde x, \tilde y$ coordinates are at distance
$\sqrt{\frac{\theta}2} \frac{\abs{\tilde z_\phi-\tilde
    z_{\phi'}}}2$. In particular the distance between the two poles
--- identified as $\omega_0, \omega_1$ --- is
$\sqrt{\frac{\theta}2}$, in agreement with Proposition
\ref{zetheorem}.
 \section{Conclusion}
\label{sec:5}
Describing the Moyal plane by a spectral triple based on the algebra
of Schwartz functions equipped with Moyal $\star$-product \eqref{eq:moyal0}, we have computed the spectral distance between the eigenfunctions of the quantum harmonic oscillator. We have shown that the distance is not finite on the whole of the pure-state space of $\A$, and that it can be made finite by truncating the Moyal spectral triple.

All the results in this paper are expressed in the matrix
  basis. It would be interesting to study the interpretation in the $x$-space.
 Can one see the pure states $\omega_n$ as deformation of some
  states of $C_0(\gR^2)$? A relevant tool on that matter could be
 the coherent states of the harmonic oscillator.

  The limit $\theta \rightarrow 0$ also deserves more attention. The pure state space for
  $\theta\neq 0$ (infinitely many pure states yielding the same
  representation modulo unitaries) is very different from the pure
  state space at $\theta=0$ (infinitely many pure states, the points
  of $\gR^2$, yielding non-equivalent representations). This question
  should be addressed by using the continuous field of $C^*$-algebras, 
$\theta\mapsto \A_\theta= \overline{(\caS, \star_\theta)}$ where $\star_\theta$
is the Moyal product for the given value of the parameter. 

  One should also question the unitization problem. What is gained,  from the metric point of view, by passing to the
  preferred unitization $\A_1$? In the commutative case, the algebra
  $C_b(X)$ of bounded continuous functions on a locally compact space $X$ is
  the maximal unitization of $C_0(X)$, and ${\cal P}(C_b(X))$ is the
  Stone-\v{C}ech compactification of ${\cal P}(C_0(X))\simeq
  X$. This is no longer true in the non-commutative
  case. In particular $\A_1$ is not the maximal unitization of $\A$
  (which is the Moyal multiplier algebra ${\cal M}$). The link between
  ${\cal P}(\A)$ and ${\cal P}(\A_1)$ deserves further studies.

Finally, in the commutative case $\A= C_0^\infty(M)$, for $M$ a complete locally compact
Riemannian manifold, the spectral distance between non-pure states
coincides with the Wasserstein distance of order $1$ between
probability distributions on $M$ \cite{DM09}.   Is there any possibility to associate a
  Wasserstein distance to the spectral distance on $S^2$ computed
  here?

\bigskip

\begin{center}{\bf Acknowledgements}\end{center}
The authors thank the anonymous referee for carefully reading
  the original manuscript and for many valuable remarks and
  suggestions. 
This work was partially supported by the ERG fellowship 237927 from the Europen Union.

\end{document}